\documentclass[twoside,twocolumn,9pt]{article}
\usepackage{extsizes}
\usepackage[super,sort&compress,comma]{natbib} 
\usepackage[version=3]{mhchem}
\usepackage[left=1.5cm, right=1.5cm, top=1.785cm, bottom=2.0cm]{geometry}
\usepackage{balance}
\usepackage{times,mathptmx}
\usepackage{sectsty}
\usepackage{graphicx} 
\usepackage{lastpage}
\usepackage[format=plain,justification=justified,singlelinecheck=false,font={stretch=1.125,small,sf},labelfont=bf,labelsep=space]{caption}
\usepackage{float}
\usepackage{fancyhdr}
\usepackage{fnpos}
\usepackage[english]{babel}
\usepackage{array}
\usepackage{droidsans}
\usepackage{charter}
\usepackage[T1]{fontenc}
\usepackage[usenames,dvipsnames]{xcolor}
\usepackage{setspace}
\usepackage[compact]{titlesec}
\usepackage{hyperref}

\definecolor{cream}{RGB}{222,217,201}

\begin{document}

\pagestyle{fancy}
\thispagestyle{plain}
\fancypagestyle{plain}{

\fancyhead[C]{\includegraphics[width=18.5cm]{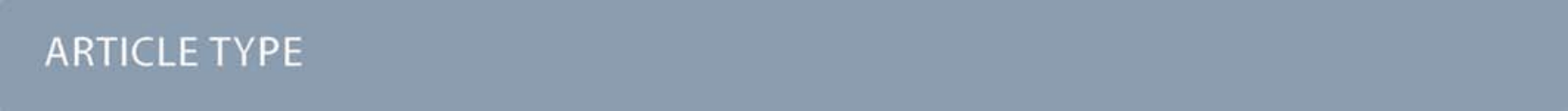}}
\fancyhead[L]{\hspace{0cm}\vspace{1.5cm}\includegraphics[height=30pt]{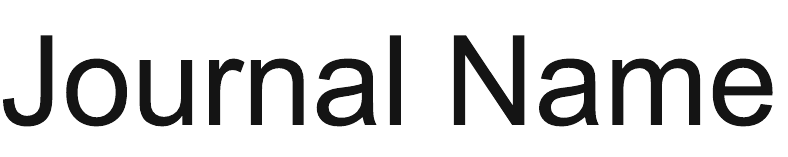}}
\fancyhead[R]{\hspace{0cm}\vspace{1.7cm}\includegraphics[height=55pt]{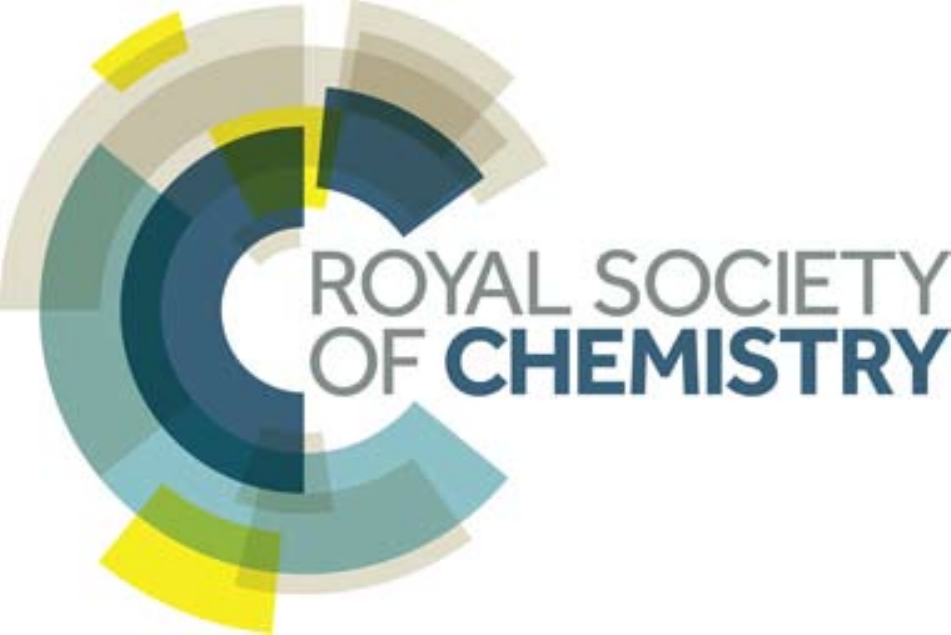}}
\renewcommand{\headrulewidth}{0pt}
}

\makeFNbottom
\makeatletter
\renewcommand\LARGE{\@setfontsize\LARGE{15pt}{17}}
\renewcommand\Large{\@setfontsize\Large{12pt}{14}}
\renewcommand\large{\@setfontsize\large{10pt}{12}}
\renewcommand\footnotesize{\@setfontsize\footnotesize{7pt}{10}}
\makeatother

\renewcommand{\thefootnote}{\fnsymbol{footnote}}
\renewcommand\footnoterule{\vspace*{1pt}%
\color{cream}\hrule width 3.5in height 0.4pt \color{black}\vspace*{5pt}} 
\setcounter{secnumdepth}{5}

\makeatletter 
\renewcommand\@biblabel[1]{#1}            
\renewcommand\@makefntext[1]%
{\noindent\makebox[0pt][r]{\@thefnmark\,}#1}
\makeatother 
\renewcommand{\figurename}{\small{Fig.}~}
\sectionfont{\sffamily\Large}
\subsectionfont{\normalsize}
\subsubsectionfont{\bf}
\setstretch{1.125} 
\setlength{\skip\footins}{0.8cm}
\setlength{\footnotesep}{0.25cm}
\setlength{\jot}{10pt}
\titlespacing*{\section}{0pt}{4pt}{4pt}
\titlespacing*{\subsection}{0pt}{15pt}{1pt}

\fancyfoot{}
\fancyfoot[LO,RE]{\vspace{-7.1pt}\includegraphics[height=9pt]{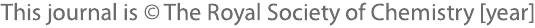}}
\fancyfoot[CO]{\vspace{-7.1pt}\hspace{13.2cm}\includegraphics{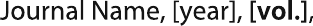}}
\fancyfoot[CE]{\vspace{-7.2pt}\hspace{-14.2cm}\includegraphics{RF}}
\fancyfoot[RO]{\footnotesize{\sffamily{1--\pageref{LastPage} ~\textbar  \hspace{2pt}\thepage}}}
\fancyfoot[LE]{\footnotesize{\sffamily{\thepage~\textbar\hspace{3.45cm} 1--\pageref{LastPage}}}}
\fancyhead{}
\renewcommand{\headrulewidth}{0pt} 
\renewcommand{\footrulewidth}{0pt}
\setlength{\arrayrulewidth}{1pt}
\setlength{\columnsep}{6.5mm}
\setlength\bibsep{1pt}

\makeatletter 
\newlength{\figrulesep} 
\setlength{\figrulesep}{0.5\textfloatsep} 

\newcommand{\topfigrule}{\vspace*{-1pt}%
\noindent{\color{cream}\rule[-\figrulesep]{\columnwidth}{1.5pt}} }

\newcommand{\botfigrule}{\vspace*{-2pt}%
\noindent{\color{cream}\rule[\figrulesep]{\columnwidth}{1.5pt}} }

\newcommand{\dblfigrule}{\vspace*{-1pt}%
\noindent{\color{cream}\rule[-\figrulesep]{\textwidth}{1.5pt}} }

\makeatother

\twocolumn[
  \begin{@twocolumnfalse}
\vspace{3cm}
\sffamily
\begin{tabular}{m{4.5cm} p{13.5cm} }

\includegraphics{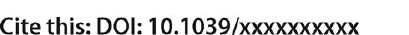} & \noindent\LARGE{\textbf{Nonaffine deformation and tunable yielding of colloidal assemblies at the air-water interface}} \\
\vspace{0.3cm} & \vspace{0.3cm} \\

 & \noindent\large{Armando Maestro$^{\ast,\ddag}$\textit{$^{a}$}  and Alessio Zaccone$^{\ast}$\textit{$^{a,b}$}} \\

\includegraphics{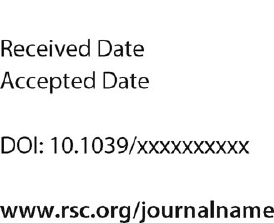} & \noindent\normalsize{Silica nanoparticles trapped at air-water interface form a 2D solid state with amorphous order. We propose a theoretical model to describe how this solid-like state deforms under a shear strain ramp up to and beyond a yielding point which leads to plastic flow. The model accounts for all the particle-level and many-body physics of the system: nonaffine displacements, local connectivity and its evolution in terms of cage-breaking, and interparticle interactions mediated by the particle chemistry and colloidal forces. The model is able to reproduce experimental data with only two non-trivial fitting parameters: the relaxation time of the cage and the viscous relaxation time. The interparticle spring constant contains information about the strength of interparticle bonding which is tuned by the amount of surfactant that renders the particles hydrophobic and mutually attractive. This framework opens up the possibility of quantitatively tuning and rationally designing the mechanical response of colloidal assemblies at the air-water interface. Also, it provides a mechanistic explanation to the observed non-monotonic dependence of yield strain on surfactant concentration. } \\

\end{tabular}

 \end{@twocolumnfalse} \vspace{0.6cm}

  ]

\renewcommand*\rmdefault{bch}\normalfont\upshape
\rmfamily
\section*{}
\vspace{-1cm}


\footnotetext{\textit{$^{a}$~Cavendish Laboratory, University of Cambridge, CB3 0HE Cambridge,
U.K.; E-mail: am2212@cam.ac.uk}}
\footnotetext{\textit{$^{b}$~Statistical Physics Group, Department of Chemical
Engineering and Biotechnology, University of Cambridge, CB3 0AS Cambridge, U.K.; E-mail: az302@cam.ac.uk}}


\footnotetext{\ddag~Present address: Institute Laue - Langevin, 71 Avenue des martyrs, 3800 Grenoble, France }


\section{Introduction}

We every day encounter materials that consist of colloidal particles adsorbed at fluid interfaces, like foams and emulsions mostly exploited in the oil-recovery industry and in food and pharmaceutical formulations.~\cite{Binks2002, Hunter2008, Calderon2008, Dickinson2010} In the last years, an increasing attention has been paid to fluid interfaces as templates for the direct assembly of inorganic nanoparticles. As a result, fluid interfaces are a versatile platform to create mechanically stable nanostructures~\cite{Russell2007, Marzan2010, KnowlesMezzenga_AM, Dai2015, Dai2017, DelGado2014} for cutting-edge applications including biosensing~\cite{Duyne2008}, and catalysis~\cite{Stellacci_2012,Mezzenga_2015} to mention a few examples. In addition, the use of ligand-nanoparticle complexes improve the control of not only adsorption to fluid interfaces but also the interfacial assembly and dynamics allowing to exploit nanoparticle interfacial layers more broadly in advanced materials applications.~\cite{Garbin_review} In these complexes, the nanoparticle core is related to the photonic or electronic properties, whereas the (physically or chemically) surface-attached ligands define the particle's adsorption and interaction in the interfacial plane.

Colloidal particles have strong affinity for fluid interfaces because the adsorption energy largely overcomes the particle's thermal energy $k_{B}T$.~\cite{Binks2002} They are also small enough (from micro- to nanoscale) to avoid gravity effects, therefore, their equilibrium position with respect to an air/water interface is determined by the balance of three interfacial energies corresponding to the air/water, particle/water and particle/air. This is accounted for by an equilibrium contact angle described by Young's equation.~\cite{Maestro_ca_2015} At a fluid interface, the particles adsorbed are highly mobile being able to achieve an equilibrium assembly that is dictated by  inter-particle interactions.~\cite{Pieranski_1980}   In general, colloidal microparticles with dissociable charged groups on their surface, repel each other due to the existence of double-layer repulsive force that counteract the attractive van der Waals attraction.~\cite{Israelachvili, Oettel2008} Additionally, capillary forces may emerge induced by local interfacial deformation between particles.~\cite{Kralchevsky2000, Oettel2005} In the case of nanoparticles, their interfacial assembly is extremely dependent on the competition between thermal fluctuations, due to the fact that nanoparticles' adsorption energies are in the order of 10-100 times larger than the thermal energy, and interfacial forces.~\cite{Bresme2007} It is in this context where a precise determination of the nanoparticles wettability is crucial to control their interfacial assembly.~\cite{Isa_CA}

We consider that it is necessary to address in the previous picture how the presence of bulk flow induces an interfacial shear deformation modifying particles' arrangement and, therefore, yielding a viscoelastic response. It has been demonstrated that particles assemble into polycrystalline~\cite{Retsch2009, Dai2012, Keim_2015, Isa_buttinoni_2017} and/or amorphous colloidal monolayers~\cite{Cicuta_superposition2003} show identical shear rheological features: under shear deformation, particle monolayers respond as linear elastic solids at small strains whereas at large enough strains, microstructural rearrangements become irreversible yielding to plastic flow behavior.~\cite{Keim_2015}

In this work, we focus our attention to the case of nanoparticle monolayers that assemble into disordered solid structures reminiscent of colloidal glasses. Several rheological studies have been performed in the last years to show the link between the deformation and flow of nanoparticle-laden interfaces with the interfacial microstructure and interparticle interactions.~\cite{Zang_2010, Liggieri_2011, Vermant_2013, Barman_2014, Maestro_Langmuir2015} In general, the presence of nanoparticles increases the rigidity of fluid interfaces and, therefore, the resistance against deformation.~\cite{Erni_review} This has been extensively exploited mainly in the stabilization of foams~\cite{Gonzenbach2006, Orsi2012, Arriaga2014, Maestro_bubbles2014} and emulsions.~\cite{Whitby2005, Monteux2007, Vermant2017}  In contrast, a fundamental explanation of the microscopic mechanism controlling the dynamics of the particles at fluid interfaces has been more elusive. 

A versatile model system  for interfacial coating consists on hydrophobic colloidal silica nanoparticles in combination with oppositely charged surfactants.~\cite{Gonzenbach2006, RaveraCSA2006, MaestroWet2012} Here, the surfactant molecules control not only the adsorption and the wettability of the particles, but also the interaction strength between particles can be tuned --and, therefore, the interfacial packing density $\phi$-- by modifying the surfactant concentration $C_{s}$; \textit{i.e.}, the amphiphilic molecules anchored at the particle surface yield to an attractive force between the nanoparticles based on the hydrophobic interaction between their hydrocarbon tails. The shear-induced deformation of this surfactant-nanoparticle complexes, at high enough surfactant concentration,  shows an overall solid-like response below a yield point with properties of a 2D glass.~\cite{Maestro_Langmuir2015} The yielding behavior of silica nanoparticles attached at air/water interfaces has been also studied by large-amplitude oscillation rheology very recently.~\cite{Harbottle_yielding_ 2016} In this work, varying both surfactant and particle concentration,  the soft-glassy dynamics is also confirmed. 

In view of the above experimental evidences, here we propose a microscopic mechanism that is responsible for the deformation of inorganic nanoparticle interfacial layers under shear. This model is based on the description of local connectivity, and its temporal dynamics, and of the microstructural heterogeneity of the elastic response (giving rise to strongly nonaffine deformations) of particles interfacial assemblies. The mathematical model is in good agreement with the oscillatory shear measurements performed providing a fundamental connection between the concept of nonaffine deformations, the dynamical rearrangements of the local cage and the onset of plastic flow - all of which can be tuned to a certain extent by means of the colloidal chemistry.

\begin{figure}[h]
\centering
\includegraphics [width=0.45\textwidth]{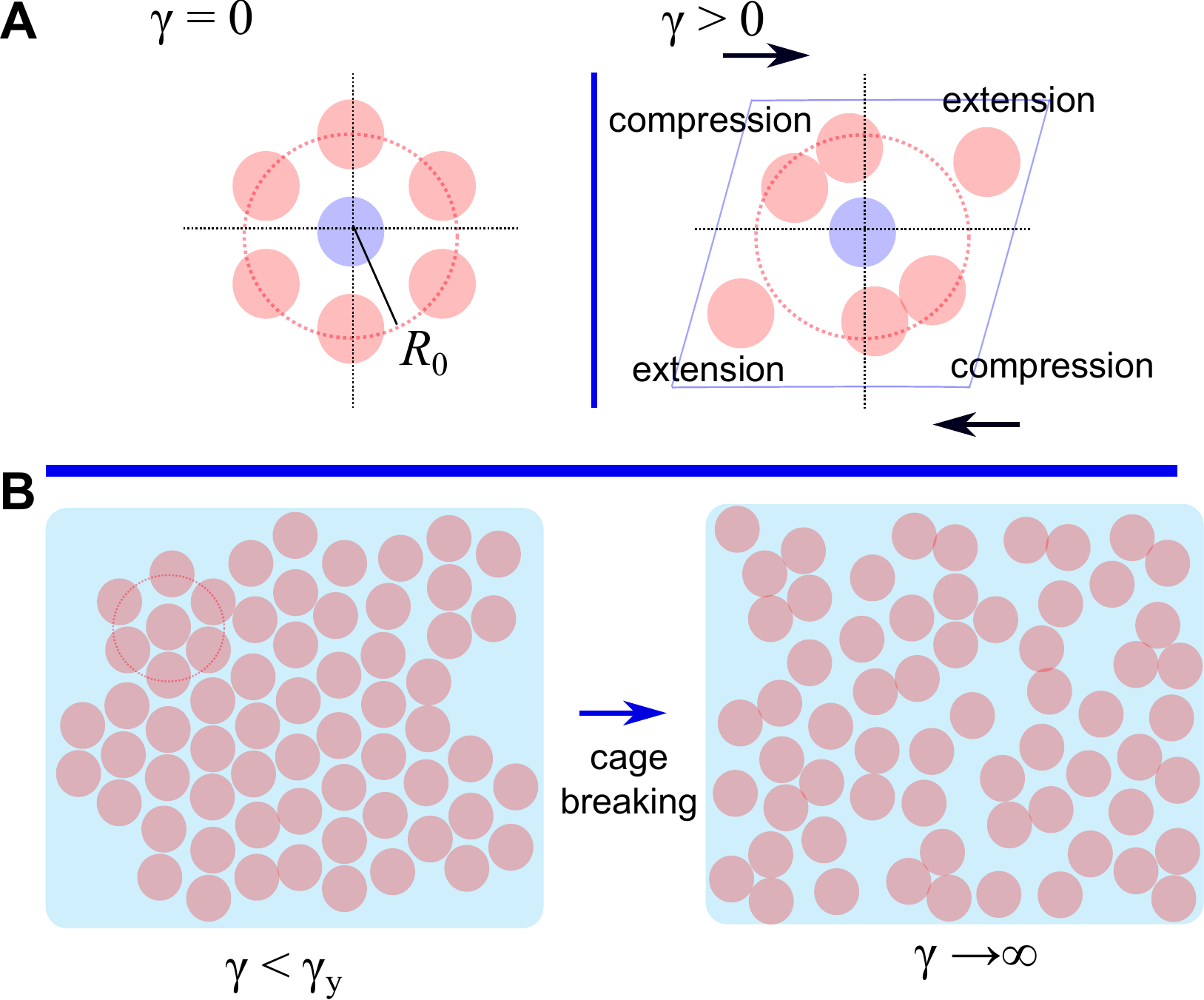}
\caption{(A) Cage-breaking model: In the absence of shear, the number of particles at the interface moving in and out of the cage is equal. In the presence of shear $\gamma$, the number of particles moving out of the cage in the sectors of the local extension axis is higher than in the sectors of the compression axis. (B) Top view of the distribution of nanoparticles at the interface in a situation of dense packing illustrating the proccess of yielding and cage breaking.}
\label{fig:scheme}
\end{figure}

\section{Nonaffine elastic deformation}
As for 3D systems,~\cite{ZacconePRB2011,ZacconePRB2014} the starting point of the analysis is the free energy of deformation of disordered solids which can be written as 
$F(\gamma)=F_{A}(\gamma) - F_{NA}(\gamma)$, with two distinct contributions arising in response to the macroscopic shear deformation $\gamma$. $F_{A}$ is the standard affine deformation energy. Affinity in this case means that every particle at the interface follows exactly the macroscopic shear deformation, and the associated interparticle displacement of a tagged particle $i$ is given by $\textbf{r}_{i}^{A}=\gamma \textbf{R}_{i}$, where $\textbf{r}_{i}^{A}$ is the affine particle position, while $\textbf{R}_{i}$ is the particle position in the rest frame. The nonaffine contribution $-F_{NA}$ lowers the free energy of deformation due to the fact that the particles at the interface are not local centers of lattice symmmetry, and, thus, there is an imbalance of forces on every particle in the affine position when the deformation is applied. This is due to the fact that the particle's nearest neighbours also react to the imposed deformation and in so doing transmit forces to the tagged particle. Clearly, these forces acting in the affine position can mutually cancel out only if the tagged particle is a center of symmetry in the lattice. In a disordered lattice, the particle is not a center of symmetry, and there is therefore a net force acting on it in the affine position.

This additional net force acting on every particle in the network has to be relaxed through additional (\textit{nonaffine}) motions that happen on top of the affine displacement dictated by the macroscopic strain. Then, the force acting on every particle times the nonaffine displacement, contributes a net work that the system has to do in order to maintain mechanical equilibrium. This work is an internal work done by the system, and defines the nonaffine contribution $-F_{NA}(\gamma)$.
In earlier work by Zaccone and Scossa-Romano ~\cite{ZacconePRB2011} it was shown that, for a disordered assembly of spheres, the resulting shear modulus is given by
\begin{eqnarray}
G&=G_{A}-G_{NA}=\frac{1}{30}\frac{N}{V}\kappa R_{0}(z-6)\\
G&=G_{A}-G_{NA}=\frac{1}{18}\frac{N}{A}\kappa R_{0}(z-4)
\end{eqnarray}
for $d=3$ and $d=2$, respectively. In general, the scaling $G\sim (z-2d)$ applies for generic $d$-dimensional systems. In the above formulae, $N/V$ and $N/A$ represent the number of particles per unit volume and per unit surface, respectively. $\kappa$ is the spring constant for nearest-neighbour interaction, defined as the second derivative of the interparticle potential evaluated at the bonding minimum. For hard-sphere systems, a spring constant can still be defined by considering the effective many-body potential from Boltzmann inversion of the radial distribution function (the effective minimum is due to entropic many-body effects). Finally, $R_{0}$ represents the equilibrium distance between its nearest neighbours and $z$ denotes the average coordination number.

\section{Cage-breaking model}
We consider here an interface covered by nanoparticles. Let us focus our attention to a single particle in the network (see Fig.~\ref{fig:scheme}A),  the presence of local shear deformation, $\gamma > 0 $, around that given particle divides the interfacial space into two different sectors of a solid angle under shear: extension and compression. In the extension sector, the neighbouring particles are pulled apart from the considered particle at the center of the cage. The neighbours cross the boundary marked by the interparticle distance $R_{0}$ in the outward direction and, thus, they do not contribute to $z$. 
In the compression sector, particles are pushed inwards by the local deformation field. However, this effect is in opposition to the existence of excluded-volume interactions between particles.
As a result, the shear-induced  depletion of mechanical bonds in the extension sectors cannot be compensated by the formation of new bonds in the compression sectors.
A simple, general expression for the evolution of the coordination number $z$  due to the thermal motion $k_{B}T$ and the shear-induced distortion of the network according to the mechanism proposed in ~\cite{ZacconePRB2014} is as follows.

The probability of finding a nearest-neighbour particle $j$ around a tagged particle $i$ at a radial distance $r$ and time $t$ is given by the van Hove space-time correlation function~\cite{Hansen}:
\begin{equation}
G(r,t)=\frac{1}{N}\langle \sum_{i=1}^{N}\sum_{j=1}^{N} \delta(\textbf{r}+\textbf{r}_{j}(0)-\textbf{r}_{i}(t))\rangle
\end{equation}
which gives the probability that two particles $i$ and $j$ are at a distance $r$ at time $t$ under the constraint that one of them was at the origin at $t=0$. 
The van Hove correlation function can be split into two contributions, the self-part $G_{s}(r,t)$ and the distinct part $G_{d}(r,t)$, respectively. The self-part represents the motion of the particle which was initially at the origin, whereas the distinct part represents the motion of the second particle relative to the first. 
The Fourier transform of the van Hove correlation function gives the intermediate scattering function which is an experimentally accessible quantity (e.g. in light and neutron scattering experiments):
\begin{equation}
F(q,t)=\int d^{3} \textbf{r} ~G(r,t)\exp(-i \textbf{q} \cdot \textbf{r})
\end{equation}
Clearly, also the intermediate scattering function can be split into a self and a distinct part, $F(q,t)=F_{s}(q,t)+F_{d}(q,t)$ which are the space-Fourier transform of $G_{s}(r,t)$ and of $G_{d}(r,t)$, respectively~\cite{Schmidt}. 

At $t=0$ the van Hove correlation function reduces to the static particle-particle autocorrelation function:
\begin{equation}
G(r,0)=\delta(\textbf{r})+\rho g(r)
\end{equation}
where the Dirac delta function comes from the self-part, while $g(r)$ is the standard radial distribution function coming from $G_{d}(r,t)$. Hence, $G_{d}(r,0)=\rho g(r)$.
Here, $\rho=N/V$ in $d=3$ or $N/A$ in $d=2$.

The static average number of nearest neighbours $Z_{0}\equiv z_{0}$ is defined in terms of the $g(r)$, as is well known, by the following relation
\begin{equation}
z_{0}=4\pi\rho\int_{0}^{R_{c}} g(r) r^{2} dr = 4\pi\int_{0}^{R_{c}} G_{d}(r,0) r^{2} dr
\end{equation}
where $R_{c}$ is a cut-off that is often set equal to the first minimum in the amorphous $g(r)$. With this choice, $z_{0}\simeq 12$ for liquids and glasses of spherical particles in $d=3$, and $z_{0}\simeq 6$ in $d=2$.

The definition of average number of nearest neighbours can be extended to the dynamic case, by replacing the static distribution function $\rho g(r)=G_{d}(r,0)$ with the time-dependent one, $G_{d}(r,t)$,
\begin{equation}
z(t)=4\pi\int_{0}^{R_{c}} G_{d}(r,t) r^{2} dr
\end{equation}
provided that a nearest-neighbour peak is identifiable also in the space-dependent part of $G_{d}(r,t)$.

At this point, in order to determine the cage dynamics, it is necessary to resort to theories of many-particle dynamics in dense liquids and glasses. Mode-coupling theory provides such a theory for the intermediate scattering function $F(q,t)$. In practice, MCT derives an equation of motion for $F(q,t)$ which is formally analogous to a generalized Langevin equation with a memory-kernel which provides a feedback mechanism to slow down the correlated particle motion~\cite{Goetze}. The final result shows that the time-decay of the intermediate scattering function for long times is dominated by the self-part and features a stretched-exponential decay~\cite{Goetze,Hansen}:
\begin{equation}
F(q,t)\sim\exp(-t/\tau_{c})^{\beta}
\end{equation}
where $\tau_{c}$ is the $\alpha$-relaxation time which is associated with substantial restructuring of the glassy cage, and the stretching exponent $\beta$ is typically in the range $0.5-0.6$~\cite{Goetze,Hansen}. In the following, we find that an excellent fit of experimental data is obtained with $\beta=0.55$ which perfectly falls within the range reported in the literature for glassy systems. 

Using the Vineyard approximation~\cite{Vineyard},
\begin{equation}
F(q,t)\simeq S(q)F_{s}(q,t)
\end{equation}
it follows that the distinct part has the same time-dependence as the self-part and the total $F(q,t)$:
\begin{equation}
F_{d}(q,t)\simeq [S(q)-1] F_{s}(q,t)\simeq \frac{[S(q)-1]}{S(q)} F(q,t).
\end{equation}
Here $S(q)$ is the static structure factor, i.e. the space Fourier transform of $g(r)$. 

Hence, assuming that $F(q,t)\sim\exp(-t/\tau_{c})^{\beta}$, we then have
\begin{equation}
G_{d}(r,t)\simeq \frac{[S(q)-1]}{S(q)}\int d^{3} \textbf{q} ~F(q,t)\exp(+i \textbf{q} \cdot \textbf{r}).
\end{equation}
Since the inverse Fourier transform over space leaves the time-dependence unaltered, hence the time-dependence of $G_{d}(r,t)$ follows as:
\begin{equation}
G_{d}(r,t)\sim \exp(-t/\tau_{c})^{\beta}
\end{equation}
and therefore also the integration over the first peak of $G_{d}(r,t)$ leaves the following dependence for the dynamic mean nearest-neighbour number:
\begin{equation}
z(t) \sim \exp(-t/\tau_{c})^{\beta}.
\end{equation}

In a liquid, the mean number of nearest-neighbours is basically constant with time. If within a time interval $\tau$ there are $z(\tau)$ neighbours that leave the first coordination shell, there are as many other particles (originally not in the first coordination shell) that replace them.
In a glass, the situation is similar although the time scale at which neighbours leave the cage is much longer.

If a glass state is put under shear, the situation becomes much different. The spherical space around a tagged particle can be subdivided into 4 quadrants (see also Fig. 1). Two of them are extensional quadrants: here particles that leave the cage, are very unlikely to get back and also other particles are very unlikely to take their places, because the field pushes them outwardly away from the tagged particle at the center of the frame. The other two quadrants are instead compressional. In these quadrants particles originally in the cage are very unlikely to leave the cage, and if the attempt to do so they are pushed back by the field inwardly towards the particle at the center of the cage. 
Hence, in a glassy state under shear there is a net loss of nearest neighbours in the extensional quadrants of the shearing field, while the number of nearest neighbours in the compression quadrants is expected to remain basically constant. 
For small particles, such that the Peclet number is very small, $Pe\ll 1$, one can assume that the escape from the cage is controlled by thermal motion rather than by shear convection. For our experiments, this is certainly the case, however for larger colloidal particles the escape mechanism could be shear-driven and convective, which would lead to larger values of $\beta$ more typical of driven systems. For example in Ref.~\cite{Laurati} where larger colloids where used, an exponent $\beta=2$ was found which is indicative of convective dynamics. 

These considerations can be combined with the above result for the time-dependence of $z$ to build a model of cage deformation breaking. We can assume that the mean number of nearest neighbours in the stable glassy assembly prior to shearing is $6$, i.e. the value in 2D that one would obtain upon integrating $g(r)$ up to the first minimum. Furthermore, based on Fig.~\ref{fig:scheme}, we can assume that particles leave the cage in the extensional direction, and only in the compression direction the number of nearest neighbours remains approximately constant. 
Upon mapping time onto strain $\gamma$ for a linear increase of strain amplitude at constant rate $\dot{\gamma}$, i.e. $\gamma = \dot{\gamma} t$, these considerations imply the following limits: $z(\gamma=0)\equiv z_{0}=6$ and $z(\gamma\rightarrow \infty)=3$. In practice this means that in the limit of infinite strain (steady-state flow), only the particles in the two compression quadrants are still next to the tagged particle at the center of the frame, as they are continuously pushed back by the flow, according to a mechanism shown already in simulations of flow of hard-sphere colloids~\cite{Brady}. 

A function which has the time dependence given by Eq. (13) and complies with the limits imposed by the shearing geometry is the following
\begin{equation}
z(\gamma)= \frac{z_{0}}{2} [1 + e^{-(A \gamma)^{\beta}}],
\label{Eq:nb}
\end{equation}
with $A = \Delta / k_{B}T + 1/ \dot{\gamma}\tau_{c}$ and we recall that $\gamma = \dot{\gamma} t$. Here $\Delta$ represents an energy barrier for the shear-induced breaking of the cage, which, in glassy systems might also be related to the glass transition temperature $T_{g}$. $\dot{\gamma}$ is the strain rate and $\tau_{c}$ is a cage relaxation time. 

\section{Nonlinear stress-strain relation}
The nonaffine free energy of deformation that takes the loss of nearest-neighbours into account due to the cage breaking effect, is written such that its second derivative gives the local shear modulus $G(\gamma)$, hence~\cite{ZacconePRB2014}:
\begin{equation}
F_{el} = \frac{1}{2}K[z(\gamma) -z_{c}] \gamma^{2}
\end{equation}
where $K \equiv(1/18)(N/A)\kappa R_{0}$ and $z_{c}=4$ for a 2D assembly. One could more formally write this nonlinear free energy of deformation using neo-Hookean models~\cite{Ogden}, but this would not change the final result and the equations that we derive in the following.

Upon inserting Eq. (14) into Eq. (15), and taking the first derivative of the free energy of deformation with respect to strain, we obtain the nonlinear stress-strain relationship for the 2D particle assembly:

\begin{equation}
\begin{split}
\sigma_{el} = \frac{\partial F_{el}}{\partial \gamma}
=  K \gamma \bigg\{2 \bigg[2 + \exp\{- [\gamma(\tilde{\Delta}+\frac{1}{\dot{\gamma}\tau_{c}})]^{\beta}\} \bigg]  -z_{c}  \bigg\} \\
 -\bigg\{\frac{1}{2} K \bigg(\tilde{\Delta}+\frac{1}{\dot{\gamma}\tau_{c}} \bigg) \exp\{- [\gamma(\tilde{\Delta}+\frac{1}{\dot{\gamma}\tau_{c}})]^{\beta}\} \gamma^{2} 
 \bigg\}
 \bigg\{\gamma [\tilde{\Delta}+\frac{1}{\dot{\gamma}\tau_{c}}]^{\beta} \bigg\}^{-1},
\end{split}
\label{Eq:sigma_elas}
\end{equation}

with $\tilde{\Delta}=\Delta / k_{B}T$ is the dimensionless energy associated with cage restructuring (which has been related to the glass transition temperature in previous work, $\tilde{\Delta}=\Delta / k_{B}T = T_{g}/T$ ~\cite{ZacconePRB2014}. The prefactor $K$ of the first term also contains the dependence of the stress-strain curve on the particle size, because $R_{0}$ is approximately equal to the particle diameter for a dense assembly. Upon considering a surface packing with fixed packing fraction $\phi=\pi(R_{0}/2)^{2} N/A$, the expression for $K$ becomes: $K=(2/9\pi)\kappa \phi/R_{0}$. Hence, as expected for a 2D solid~\cite{ZacconePRB2011}, the initial slope of the linear elastic regime (hence the shear modulus) of the stress-strain relation decreases upon increasing the particle size as $K \sim R_{0}^{-1}$. In 3D we would have $K \sim R_{0}^{-2}$, and in general $K \sim R_{0}^{1-d}$ in a generic space dimension $d$. Furthermore, $K$ also appears in the first bracket of the second negative term in Eq. (16), but it could be collected as a common factor in front of all brackets and hence does not affect the position of the yielding point (which is controlled by the expressions inside the brackets and by the competition between positive and negative terms therein). Since the first bracket in Eq. (16) is the one which controls the linear elastic regime, the elastic rigidity is therefore inversely proportional to $R_{0}$.

Beside the elastic contribution, we also need to consider the dissipative contribution, $\sigma_{v}$, to the total stress. It is known that for deformations that are not quasi-static, i.e., with $\dot{\gamma} > 0$, microscopic friction induces a resistance to the particle displacements. This friction is associated with a viscosity $\eta$ and a viscous relaxation time $\tau_{v}$ according to the Maxwellian viscoelastic model~\cite{Hansen}.The viscous stress is then defined as
\begin{equation}
\sigma_{v} = \dot{\gamma} \eta\bigg[1 - \exp\bigg(-\frac{\gamma}{\dot{\gamma}\tau_{v}}\bigg)\bigg].
\label{Eq:sigma_vis}
\end{equation}

Finally, the total stress is the sum of the elastic (Eq.~\ref{Eq:sigma_elas}) and the viscous Eq.~\ref{Eq:sigma_vis} stress contributions~\cite{LandauLifshitz}
\begin{equation}
\sigma = \sigma_{el}(\gamma) + \sigma_{v}(\gamma),
\label{Eq:sigma}
\end{equation}
where $\sigma_{el}(\gamma)$ is given by Eq. (16) while $\sigma_{v}(\gamma)$ is given by Eq. (17).

This equation contains all the relevant particle-level physics: interparticle potential (contained in $K$), nonaffine displacements (associated to $z_{c}$ and contained in Eq.(16), shear-induced changes in the local particle network $z(\gamma)$, and also includes the thermally activated cage distortion, and the viscous dissipation due to microscopic friction. The equation recovers the elastic limit at small strain, where $\sigma \approx K (z_{0}-z_c) \gamma $, and the plastic flow $\sigma \rightarrow \eta \dot{\gamma} $ in the limit $\gamma \gg 1$.

\section{Comparison}
The theory has been tested on interfacial layers consisting of the model system previously introduced: hydrophilic colloidal silica nanoparticles in combination with oppositely charged surfactants adsorbed at air/water interfaces. In an earlier work, Maestro \textit{et al} demonstrated that the strength of interaction between neighbouring nanoparticles and, therefore, their interfacial network strongly depends on the concentration of surfactant.~\cite{Maestro_Langmuir2015} In principle, the interfacial assembly of silica nanoparticles ($15\,$nm radius) is dominated by van der Waals attractive forces, and electrostatic double-layer repulsive force between the particles (because silica nanoparticles have dissociable silanol groups on their surface) according to the DLVO theory.~\cite{DLVO1, DLVO2} In addition, a short-range attraction between the nanoparticles induced by the surfactant molecules anchored at the particle surface is present. This is known as an hydrophobic interaction between the hydrocarbon tails of CTAB molecules that likely dominates at short distances (in the range of surfactant chain length)~\cite{Israelachvili2011}. As a result, the number density of nanoparticles at the interface increases with the amount of surfactant used. The oscillatory interfacial rheology measurements performed in~\cite{Maestro_Langmuir2015} showed the existence of a solid-like behavior below a yield point for all the samples with a concentration of surfactant below the CMC. This network of interconnected CTAB-Silica complexes can be, thus, rationalised as an attractive glass with the yield stress scaling with the range of attraction (i.e., the surfactant concentration.) 

\begin{figure}[h]
\centering
\includegraphics [width=0.45\textwidth]{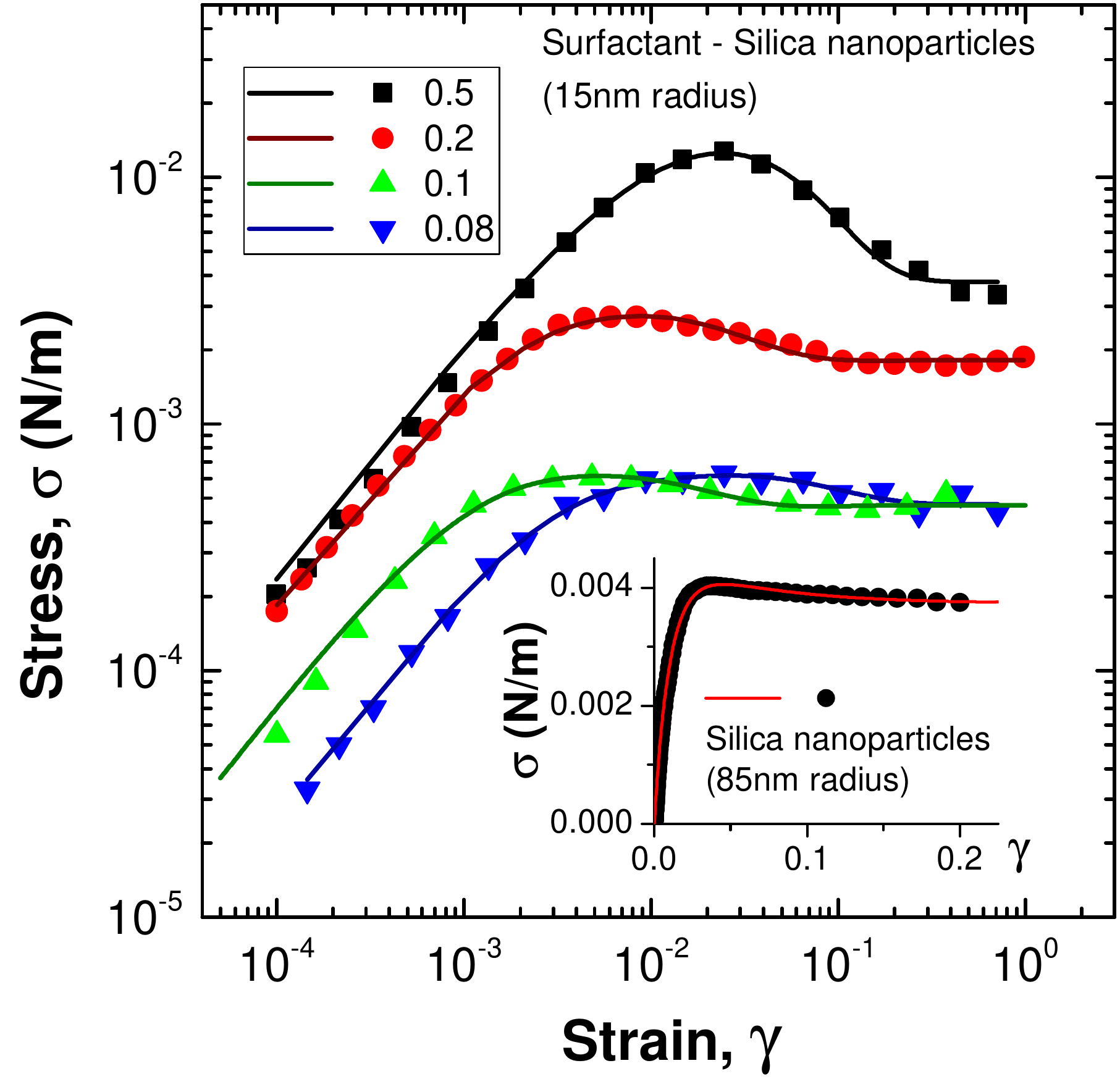}
\caption{Comparison between the theoretical expression and the experimental results by oscillatory interfacial shear rheometry obtained from~\cite{Maestro_Langmuir2015}.  Interfacial shear stress $\sigma$ is plotted versus the amplitude of strain $\gamma$ at a constant frequency $\omega=0.628\,$rad/s considering different concentrations of surfactant $C_{s}$ (expressed in mM). Eq.~\ref{Eq:sigma} is used with the viscosity $\eta$ and the strain rate $\dot{\gamma}$ fixed by the experiments. The values of $\tilde{\Delta}=6$, $\alpha=0.55$ and $z_{0}=6$ has been chosen accordingly to represent a disordered, amorphous solid network at the interface. Inset: Comparison between the theory (Eq.~\ref{Eq:sigma}) and the experimental $\sigma - \gamma$ for partially hydrophobic silica nanoparticle monolayers  ($85\,$nm radius) obtained from~\cite{Zang_2010}. In this case, $\eta$ and $\dot{\gamma}$ were also fixed by the experiments and the values of $\alpha$, $\tilde{\Delta}$ and $n_{b}^{0}$ are similar than in the main figure. 
}
\label{fig:results}
\end{figure}

In particular,  the shear stress $\sigma$ of the interfacial film at constant frequency $\omega$ (0.628$\,$rad/s), was measured in~\cite{Maestro_Langmuir2015}  varying the strain amplitude $\gamma$. Fig.~\ref{fig:results} shows the experimental strain-sweep experiments performed in~\cite{Maestro_Langmuir2015} for samples at increasing CTAB concentration $C_{s}$. In all the samples studied, the stress $\sigma$ linearly grows with $\gamma$ at low values of $\gamma$ below a certain threshold known as the limit of linearity $\gamma_{0}$. This marks the end of the linear regime, with $\sigma \propto \gamma$. Beyond such regime, there is a non-linear regime where $\sigma$ increases sub-linearly until the local stress is maximum $\sigma_{y}$ at the yield-point $\gamma_{y}$. Beyond this point, the observed behaviour depends strongly on $C_{s}$. At low $C_{s}$, the yield-point is followed by a plastic flow regime characterised by a practically constant plateau stress; i.e.,  the sample is shear-melted with $\sigma \propto \dot{\gamma} \sim \omega \gamma $. 
Upon increasing $C_{s}$ and beyond $\gamma_{y}$, the stress progressively falls down with the strain. In general, the stress-strain relation shows an overshoot with a maximum in the stress beyond which the system yields a viscous Newtonian flow in the large strain limit, $\gamma \gg 1$. This behavior, illustrated in Fig.~\ref{fig:results}, can be qualitatively rationalised as a weakly attractive glass that exhibits one-step yielding at increasing oscillatory strain amplitude. The stress overshoot is a hallmark of the rheology of glassy materials, and is observed for example in colloidal glasses~\cite{Poon_yielding, Ballauff, Harbottle_yielding_ 2016, attractive_glass} as well as in metallic glasses.~\cite{Johnson,ZacconePRB2014} 
This behavior, which is well known in 3D glasses, was indeed found in other nanoparticle monolayers consisting on partially hydrophobic silica with larger size ($85\,$nm radius) in absence of surfactant. In this case, the partial hydrophobicity of the particles is obtained by replacing the silanol by silane groups at the surface of the particles. This data, obtained from~\cite{Zang_2010} has been also plotted as an inset in Fig.~\ref{fig:results} to be compared with the proposed theory.

We compare here this nonlinear stress-strain behaviour found experimentally in 2D nanoparticle systems with a description based on the coupling between many-body dynamics causing structural rearrangments of the glassy cage and a nonaffine response to deformation. The resulting fully analytical expression for the stress-strain curve (Eq.~\ref{Eq:sigma}, together with ~\ref{Eq:sigma_elas} and ~\ref{Eq:sigma_vis}) is in good agreement with the experiments being able to reproduce the limit of linearity, the strain-softening, the yielding point and the stress overshoot upon increasing $C_{s}$ as it can be seen in Fig.~\ref{fig:results}. There is remarkable agreement between the theoretical expression (Eq.~\ref{Eq:sigma}) and the experimental data across a broad range of parameters. The values for the viscosity $\eta$, the strain rate $\dot{\gamma}$ are fixed by the experimental values obtained from Ref~\cite{Maestro_Langmuir2015} and shown in Table~\ref{tab:results}.  We fixed also a value of $\tilde{\Delta}=\frac{T_{g}}{T} = 6$ corresponding to about $1\,k_{B}T$ of attraction per nearest-neighbour as expected for weak attraction, and a value of $\alpha=0.55$, typical of glassy systems~\cite{Goetze,Hansen} for the stretched exponential of $\alpha$-relaxation~\cite{Goetze}. The only parameters that have not been fixed in Eq.~\ref{Eq:sigma} to fit the experimental data are the 'dressed' spring constant parameter $K$ (related to the interparticle potential via the spring constant $\kappa$ as mentioned earlier) and the two relevant relaxation times $\tau_{c}$, for the cage structural rearrangement and $\tau_{v}$, for the macroscopic viscous relaxation, respectively.  

Furthermore, to show the universality of the theoretical model proposed, we also compare it  with the stress-strain behavior corresponding to larger, partially hydrophobic nanoparticles adsorbed at the air/water interface in the inset of Fig.~\ref{fig:results}. In this case, there is also a good description of the experimental $\sigma-\gamma$ data with Eq.~\ref{Eq:sigma} using the values of  $\alpha$, $\tilde{\Delta}$, $z_0$ and $z_c$ previously described for glassy systems of spherical building blocks with central-force interaction. The values obtained from the fit $K=0.06\,$N/m, $\tau_{c} = 152\,$s and $\tau_{v}=5.6\times10^{-3}\,$s are in good agreement with those corresponding to the surfactant-decorated silica nanoparticles with smaller size shown in Table 1. In detail, the spring constant $K$ is in the same order than the CTAB-decorated silica nanoparticles with $0.1-0.08\,$mM CTAB as expected for a similar packing fraction of the particles at the interface. The value of $\tau_{c}$ is bigger than the ones shown in Table 1 which is physically meaningful because of the increase in particle size that, therefore, increases the Brownian relaxation time and hence $\tau_{c}$ as well.

\begin{figure}[h]
\centering
\includegraphics [width=0.6\textwidth]{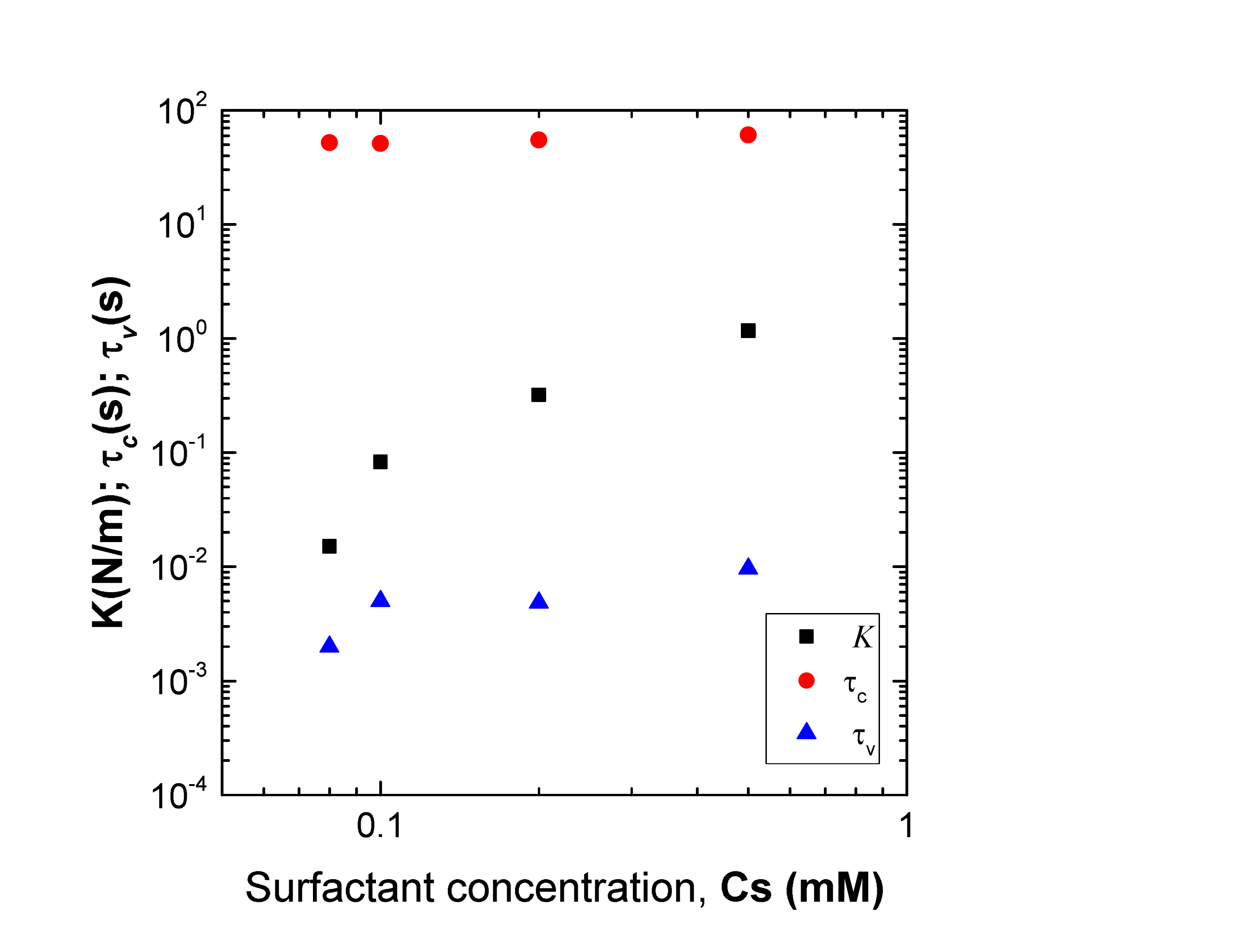}
\caption{Dependence of the fitting parameters obtained with the concentration of surfactant used to increase the hydrophobicity of the nanoparticles and, therefore, the packing density at the interface.}
\label{fig:fit}
\end{figure}

\section{Effect of surfactant on the stress overshoot}
Looking at the experimental data in Fig.~\ref{fig:results}, one can see that the amplitude of the overshoot represented by $\sigma_{y}$, and also the yield strain $\gamma_{y}$, depend on the surfactant concentration $C_{s}$. In general terms, the model explains the existence and the amplitude of the overshoot based on the competition between the elastic instability driven by non-affine cage breakup and the build-up of viscous stress, respectively. When the elastic instability sets in, it causes the stress to go through a maximum value $\sigma_{y}$ and to subsequently decrease with further increasing strain, whereas the viscous contribution $\sigma_{v}$ increases monotonically up to the final Newtonian-like viscous plateau where $\sigma \sim \dot{\gamma}\eta$. 

Fig.~\ref{fig:fit} (and Table~\ref{tab:results}) shows the dependence of the parameters $K$, $\tau_{c}$ and $\tau_{v}$ on the concentration of surfactant $C_{s}$, and, therefore, on the particle density at the interface $\phi=f(C_{s}, \,C_{p})$ --being the concentration of particles $C_{p}$ fixed in all the cases to $1 wt.\%$--. Remarkably, the prefactor $K$ of the elastic free energy, defined as $K=(2/9\pi)\kappa\phi/R_{0}$ as discussed above,  increases with $C_{s}$. This is a reasonable outcome because 
the increase of surfactant brings about an increase of the packing fraction $\phi$ of the particles at the interface~\cite{Maestro_Langmuir2015}, and at the same time an increase of the spring constant $\kappa$. The latter is defined as the second derivative of the total interaction energy between two particles evaluated in the attractive minimum. This quantity is expected to increase as a result of the increased attractive force due to hydrophobicity, and because the attractive minimum becomes narrower as the first neighbours distance $R_{0}$ decreases when more surfactant is added to the system.  

Upon increasing $C_{s}$, the surfactant-particle complexes are progressively creating a denser and stronger particle cage that results in the strengthening of the glassy network in which more energy is needed for the particles to escape from the cage, which is primarily related to $\sigma_{y}$. 
$\tau_{c}$ represents the cage relaxation time and it slightly increases with $C_{s}$. This means that the cage dynamics becomes slower upon strengthening the particle network.
Finally, we rationalise the increase of $\tau_{v}$ as due to the increase of the microscopic friction in between the particles. This friction increases as the nearest-neighbour distance decreases upon increasing the surfactant. In particular, if we visualize the nanoparticles as surrounded by a shell of vertically oriented surfactant molecules, it is clear that the friction must increase markedly when the particles approach the distance of close contact between the respective surfactant layers. This point and its consequences on the rheology are explored in the paragraph below.  

\section{Nonmonotonic dependence of yield strain on surfactant concentration}
As is clear from Fig.~\ref{fig:results}, there is a non-monotonic dependence of the yield-strain amplitude (\textit{i.e.} evaluated at the point of maximum of the stress-strain overshoot) $\gamma_{y}$ and the surfactant concentration $C_{s}$. 

In particular, $\gamma_{y}$ decreases with surfactant concentration upon going from $C_{s}=0.08$ to $C_{s}=0.1\,$mM, after which, however, it monotonically increases upon further increasing $C_{s}$. This non-monotonic behaviour would be impossible to explain without a model, but thanks to the theoretical fitting shown above, we can provide a possible physical explanation for this effect. As is shown in Fig.~\ref{fig:fit}, the time-scale associated with viscous friction, $\tau_{v}$, increases markedly upon going from $C_{s}=0.08$ to $C_{s}=0.1\,$mM, after which it is practically constant. 
As discussed above, the friction time scale $\tau_{v}$ is due to the local frictional interaction between layers of surfactants on nearest-neighbour particles. If the particles are sufficiently far apart, there is little interaction between the layers and therefore also the viscous time $\tau_{v}$ is lower. As soon as the particles become closer to each other and the layers start to interact, it is expected that the friction time scale $\tau_{v}$ increases significantly. Beyond this point, particles are unlikely to come closer due to strong steric repulsion, upon further increasing $C_{s}$. Hence, we can speculate that, upon going from $C_{s}=0.08$ to $C_{s}=0.1\,$mM, the nearest-neighbour particles become close enough such that the respective surfactant layers start to interact, which generates viscous friction in the relative motion between the particles. This would perfectly explain the jump in $\tau_{v}$ in Fig.~\ref{fig:fit} 
upon going from $C_{s}=0.08$ to $C_{s}=0.1\,$mM. Upon increasing $C_s$ further above $C_{s}=0.1\,$mM, the distance between nearest-neighbours cannot decrease further because of the strong steric repulsion between surfactant chains protruding in the layers. Hence, also the frictional time $\tau_{v}$ increases much less at this point, and saturates to a plateau. 

With reference to the non-monotonic dependence of $\sigma_{y}$ on $\gamma$, the sharp increase of $\tau_{v}$ implies that the build-up of dissipative stress $\sigma_{v}$, according to Eq.(17), becomes much slower with increasing $\gamma$. Hence the overshoot must happen at a lower strain, because the drop of $\sigma_{el}$ (which is a function featuring a maximum, that shifts towards larger $\gamma$ upon increasing $K$) is not compensated by a sufficiently fast increase of $\sigma_{v}$, as one goes from $C_{s}=0.08$ to $C_{s}=0.1\,$mM. Therefore, this consideration suggests that the drop of $\gamma_{y}$ upon going from $C_{s}=0.08\,$mM to $C_{s}=0.1\,$mM can be due to the fact that  whilst the elastic stress increases (at a given $\gamma$) due to the increase of $K$, yet the stress overshoot happens "earlier" (at a lower $\gamma$) because the viscous stress does not catch up fast enough to compensate the drop of $\sigma_{el}$, due to a larger value of $\tau_v$ which makes the increase of $\sigma_v$ with $\gamma$ much slower.

Finally, at $C_s > 0.1\,$mM the $\tau_{v}$ is basically constant, while $K$ keeps increasing. This means that the drop of elastic stress will occur at increasingly higher strain (because the maximum in $\sigma_{el}$ gets shifted to larger $\gamma$ upon increasing $K$), which is reflected in $\gamma_{y}$ increasing monotonically with $C_{s}$ in this regime.

\begin{table}[h]
\small
\centering
\caption{Relevant parameters in Eq~\ref{Eq:sigma}. The effective viscosity of the layer $\eta$ and the strain rate $\dot{\gamma}$ values are fixed by the experimental values obtained from Ref.~\cite{Maestro_Langmuir2015}. The values of the elastic constant $K$,  (related to the interparticle potential) and the two relevant relaxation times $\tau_{c}$, for the cage rearrangement and $\tau_{v}$, for the macroscopic viscous relaxation have been obtained from the fitting of the experimental data by Eq~\ref{Eq:sigma}.
 \label{tab:results}}
\begin{tabular}{l*{5}{c}r}
CTAB (mM)             & K (N/m) & $\tau_{c} (s)$ & $\tau_{v} (s)$ & $\eta$ (Ns/m) & $\dot{\gamma}$ (s$^{-1}$) \\
\hline
0.5 			& 1.17  	& 61 	&  1$\times10^{-2}$ 	& 4.9$\times10^{-3}$	& 0.82  \\
0.2          & 0.32  	& 58		&  5$\times10^{-3}$	& 5.8$\times10^{-3}$ 	& 0.28 \\
0.1          & 0.083 	& 51 	&  5$\times10^{-3}$ 	& 3.0$\times10^{-3}$	& 0.15  \\
0.08     	& 0.015	& 52		&  2$\times10^{-3}$  & 3.2$\times10^{-3}$	& 0.12  \\
\end{tabular}

\end{table}

\section{Conclusions}
We have proposed a microscopic mechanism that explains the deformation of surfactant-decorated silica nanoparticle interfacial layers under shear deformation. Silica particles trapped at the air/water interface forms a 2D amorphous solid with features of a colloidal glass. The theoretical model proposed has been able to describe how this solid-like system deforms under a shear strain ramp up and beyond a yielding point which leads to plastic flow. In detail, the model is based on the description of nanoparticle interfacial assemblies by means of the local connectivity between particles, and its temporal dynamics, and of the microstructural heterogeneity of the elastic response giving rise to strongly nonaffine deformations. 

The model is able to reproduce experimental data from oscillatory shear measurements with only two non-trivial fitting parameters: the relaxation time of the cage and the viscous relaxation time. The interparticle spring constant contains information about the strength of interparticle bonding which is tuned by the amount of surfactant that renders the particles hydrophobic and mutually attractive. This model, therefore, shows --for the first time in interfacial systems-- a fundamental connection between the concept of nonaffine deformations, the dynamical rearrangements of the local cage and the onset of plastic flow --all of which can be controlled to a certain extent by the surfactant and particle concentration--.

Finally, this framework opens up the possibility of quantitatively tuning and rationally designing the mechanical response of colloidal assemblies at the air-water interface as it can be stated from the goodness of the theoretical model for different nanoparticle systems compared in this study. We are now  extending our theoretical description to nanoparticle systems at fluid interfaces taking into account not only the number density but also to explore in detail the particle's size effects (from the nm to the $\mu$m range) and also particle's shape (by studying different geometries like cylindrical, ellipsoidal and 'Janus') and surface roughness.

\section{Acknowledgments}
We are grateful to D. Langevin, P. Cicuta and Thomas Voigtmann for very useful discussions. AM acknowledge funding from a Royal Society Netwon International Fellowship. We are grateful to D. Langevin and D. Zang for sharing their interfacial rheological data on silica nanoparticles. 
\section*{Conflict of interest}
There are no conflicts to declare.


\balance



\bibliographystyle{rsc} 

\end{document}